\documentclass[runningheads]{llncs}
\usepackage[T1]{fontenc}
\usepackage{algorithm}
\usepackage{algorithmic}
\usepackage{xcolor}
\usepackage[most]{tcolorbox}  
\usepackage[square,sort,comma,numbers]{natbib}

\usepackage{breakurl}

\begin{document}
\title{AI based Multiagent Approach for Requirements Elicitation and Analysis}
%
%
\author{
    Malik Abdul Sami\inst{1}\orcidID{0000-0001-5136-2587} \and  
    Muhammad Waseem\inst{2}\orcidID{0000-0001-7488-2577} \and
    Zheying Zhang\inst{1}\orcidID{0000-0002-6205-4210} \and
    Zeeshan Rasheed\inst{1}\orcidID{0000-0001-9655-3096} \and
    Kari Systä\inst{1}\orcidID{0000-0001-7371-0773} \and
    Pekka Abrahamsson\inst{1}\orcidID{0000-0002-4360-2226}
}

\institute{
    Tampere University, Tampere, Finland \\
    \email{malik.sami@tuni.fi, zheying.zhang@tuni.fi, zeeshan.rasheed@tuni.fi, kari.systa@tuni.fi, pekka.abrahamsson@tuni.fi} \and
    University of Jyväskylä, Jyväskylä, Finland \\
    \email{muhammad.m.waseem@jyu.fi}
}

\maketitle              

\begin{abstract}
Requirements Engineering (RE) plays a pivotal role in software development, encompassing tasks such as requirements elicitation, analysis, specification, and change management. Despite its critical importance, RE faces challenges including communication complexities, early-stage uncertainties, and accurate resource estimation. This study empirically investigates the effectiveness of utilizing Large Language Models (LLMs) to automate requirements analysis tasks. We implemented a multi-agent system that deploys AI models as agents to generate user stories from initial requirements, assess and improve their quality, and prioritize them using a selected technique. In our implementation, we deployed four models, namely GPT-3.5, GPT-4 Omni, LLaMA3-70, and Mixtral-8B, and conducted experiments to analyze requirements on four real-world projects. We evaluated the results by analyzing the semantic similarity and API performance of different models, as well as their effectiveness and efficiency in requirements analysis, gathering users' feedback on their experiences. Preliminary results indicate notable variations in task completion among the models. Mixtral-8B provided the quickest responses, while GPT-3.5 performed exceptionally well when processing complex user stories with a higher similarity score, demonstrating its capability in deriving accurate user stories from project descriptions. Feedback and suggestions from the four project members further corroborate the effectiveness of LLMs in improving and streamlining RE phases.

\end{abstract}


\section{Introduction}

In agile software projects, requirements are often expressed as user stories \cite{userstoriejune}, which are brief descriptions of functionalities or features from the user’s perspective, emphasizing their needs and the value the feature brings. Documenting user stories is a continuous process in agile software projects that ensures the communication of user needs and technical details to developers in a non-technical language \cite{herwanto2024automating}. Writing good user stories is essential in software projects, as they convey the
needs and perspectives of users and guide the development team in implementing the expected functionalities. Poorly specified user stories can be exhaustive, biased, prone to human error, and often lack quality \cite{alawaji2024evaluating}. These challenges demand the precise conversion of user requirements into technical specifications efficiently and effectively to meet user expectations accurately.

Recently, advances in natural language processing (NLP) have paved the way for transformative applications in various domains, such as healthcare, cybersecurity, e-commerce, social media, IoT, and more \cite{xie2023chatunitest}. Among these developments, the advent of large language models (LLMs) has emerged as a powerful tool with the potential to revolutionize software engineering practices, particularly in software requirements engineering tasks. Here lies the potential for LLMs to expedite and streamline the requirements analysis process by providing efficient and effective support for requirement elicitation, documentation, and prioritization.

Additionally, using LLMs as agents have shown substantial progress in addressing complex software development challenges by simulating various development roles \cite{zzyllm, rasheed2024codepori, feng2024prompting, waseem2023artificial}. Various studies have investigated multi-agent integration within LLM frameworks \cite{zhou2023language}, focusing on software design and code generation within the Software Development Life Cycle (SDLC). Although \cite{zzyllm} reported initial findings on the use of LLMs as agents for enhancing user story qualities, studies on integrating multi-agent LLMs into the RE process, particularly generating and managing user stories, have been rarely addressed. 
Although extensive research exists on the use of LLMs as agents to assist a specific requirements task, there is a lack of tools that involve multiple stakeholders, such as technical managers, product owners, quality assurance specialists, and developers, for requirements analysis, including the generation and prioritization of user stories. Consequently, we formulated two research questions (RQs), as follows.
\begin{itemize}
\item RQ1. How can AI agents assist in requirements analysis?
\item RQ2. How effectively do different language models perform in requirements analysis tasks?
\end{itemize}

%


 To answer RQ1, we have developed a multi-agent tool that generates user stories. We use two user story quality frameworks to assess and improve the generated user stories. We also implement three prioritization techniques to prioritize user stories. To answer RQ2, we conducted several experiments and empirical studies on four project descriptions. We analyzed the results using different metrics, such as API performance and similarity scores. Additionally, we counted unique epics, the distinct number of user stories, and the word count of LLMs in requirements engineering tasks. This exploration aims to uncover the underlying reasons for effectiveness variations across these tasks.

The rest of this paper is structured as follows. Section \ref{GenerativeAI} provides information about Generative AI in software development, and Section \ref{methodology} describes the study methodology. In Section \ref{experiment}, we discuss the experiment design, and the results of this study are presented in Section \ref{results}. Section \ref{discussion} provides the discussion of the study, and Section \ref{conclusion} provides the conclusion.

\section{Generative AI in Software Engineering}
\label{GenerativeAI}

Generative AI, supporting models such as Generative Pre-trained Transformers (GPT) and Generative Adversarial Networks (GANs), produces human-like outputs.
This technology enhances text production, machine translation, conversational systems, and code generation through transformer architectures that recognize contextual relationships in text \cite{sami2024prioritizing, sami2024system}. Recent advancements highlight GPT models' potential to revolutionize software engineering by automating tasks in the SDLC, including error identification, code generation, and documentation \cite{feng2023investigating, treude2023navigating}. Integrating GPT models into software development improves efficiency and quality, accelerating the development process \cite{dong2023self, ma2023scope}.

Relevant documentation, crucial for reducing development costs, benefits from powerful LLMs and agent-based techniques. These methods define roles and facilitate communication among LLM agents, sometimes utilizing external tools. For instance, a test executor agent provides test logs via a Python interpreter \cite{huang2023agentcoder}, while a debugger agent generates control flow graphs for flaw detection \cite{zhong2024ldb}. Studies simulate human roles for LLMs, addressing autonomous development challenges with multi-AI agents working on system design, code review, and test engineering \cite{rasheed2024codepori}. LLMs are used throughout the SDLC for requirements gathering, code generation, and testing, creating a unified, efficient platform \cite{rajbhoj2024accelerating}. In particular, Zhang et al. \cite{zzyllm} presented their early findings using LLMs as agents in requirements analysis to enhance user story clarity, comprehensibility, and its alignment with business objectives.

Despite the rapid advancements in applying Generative AI models to support requirements tasks in software projects, concerns about AI hallucination remain, as it can lead to plausible yet inaccurate or irrelevant responses. Addressing how to equip AI agents with comprehensive domain knowledge necessitates further research \cite{sauvola2024future}. Effective solutions must understand diverse client needs, generate user stories, create UML diagrams, and efficiently organize test cases \cite{nguyen2023generative}. Integrating an LLM-supported web-based platform presents new research opportunities by generating user stories, assessing their quality, enabling multi-agent prioritization, and gathering stakeholder feedback.

\section{Proposing Agents for Requirements Analysis}
\label{methodology}
\begin{figure}
\includegraphics[width=\textwidth]{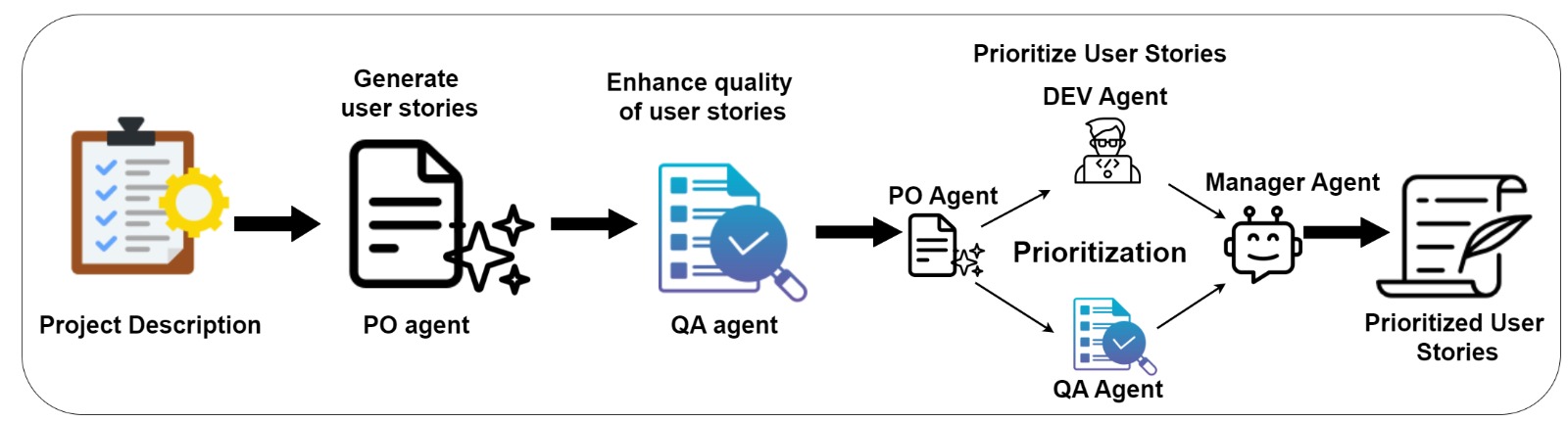}
\caption{Process Model of LLM Agents for User Story Generation, check the quality of the story, and Prioritization.} 
\label{figure1}
\end{figure}

To address the RQs, we design a multi-agent system, using LLMs as agents, to support requirements analysis tasks. This section provides an overview of the engagement of agents mainly in 
three tasks, i.e. generation of user stories based on initial project description, enhancement of user story, and prioritization of user stories. To illustrate our approach, we draw a process model that is referred to in Figure \ref{figure1}.

We designed four agents for this process: the Product Owner (PO), Developer (Dev), Quality Assurance (QA), and Manager agents. The PO is responsible for both generating and prioritizing user stories. The QA agent assesses the quality of user stories and contributes to their prioritization based on their understanding of requirements. The Developer suggests prioritizing stories from a technical feasibility perspective. The LLM Manager synthesizes suggestions from different agents to finalize the prioritization after discussions with the PO, QA, and Developer. Table \ref{tab1} describes every agent's tasks and responsibilities.

\begin{table}[ht]
\caption{Main tasks, prompts, and summary of the four agents}
\label{tab1}
\resizebox{\textwidth}{!}{%
\begin{tabular}{|l|p{2.5cm}|p{3.5cm}|p{8cm}|}
\hline
\textbf{No} & \textbf{Agent} & \textbf{Main Work} & \textbf{Prompts Summary} \\ \hline
1 & Product Owner Agent & User Story Generation and Communication, Prioritization of Stories & This prompt guides an assistant in creating detailed user stories and corresponding epics from project descriptions, focusing on unique, functional, and technical aspects to enhance project organization and implementation efficiency. It also outlines dialogue among stakeholders, including the Product Owner, to discuss and prioritize these user stories in their backlog using specified techniques. \\ \hline
2 & Quality Assurance Agent & Quality Assurance Assessment and Prioritization of Stories & This prompt instructs an assistant to evaluate user stories against a specified framework, assessing compliance, identifying issues, and documenting details including the story's description, status, and associated epic. It also ensures the prioritization of requirements by understanding the context and ensuring coverage and effectiveness. \\ \hline
3 & Sr Developer Agent & Prioritizing User Stories & Acts as a senior developer to understand project requirements and suggest the prioritization of the requirements by understanding the context. \\ \hline
4 & LLM-Manager Agent & Combining and Ranking Final Prioritization & This prompt instructs an assistant to prioritize user stories using a specified prioritization technique—such as WSJF, 100 dollars, and AHP—based on the provided context and previous discussions. The assistant is tasked with allocating points or prioritizing features to reflect their relative importance, ensuring that the distribution or ranking adheres to the required format and is supported by detailed justifications. The LLM Manager merges the prioritization from the other agents to ensure the best prioritization results. \\ \hline
\end{tabular}
}
\end{table}

\subsection{User Story Generation and Quality Assessment Agent}

The PO agent is responsible for crafting user stories based on a comprehensive understanding of the project's requirements and user needs. This process adheres to the organization's standard for documenting user stories that include user story title, description, 
and acceptance criteria. The story description clarifies the user, activity, and goal to achieve.  Subsequently, the QA agent validates  
user stories 
to assess their quality, ensuring they meet the predefined standards.

\subsection{Multiagent Agents for Prioritization}
The four agents, i.e. PO, Dev, QA, and Manager, collaborate to prioritize requirements. Like the practice in a real-life project, the PO agent initiates the process by introducing the user stories and organizing a meeting. This meeting begins with a greeting and an overview of the user stories that need prioritization. During the meeting, the Dev and QA agents receive instructions on the prioritization techniques. 
They discuss the importance and urgency of each user story and work simultaneously to prioritize them. 

Their prioritization results are then forwarded to the LLM-based Manager agent, which compiles the inputs and generates the final prioritization results, ranking the user stories according to the selected techniques. The final prioritized requirements are
exported in CSV format for 
documentation.

\section{Experimental Design}
\label{experiment}
In this study, we utilized established guidelines (e.g., \cite{wohlin2012experimentation}, \cite{runeson2009guidelines}) to design and conduct experiments in software engineering to ensure a rigorous, structured, and reproducible approach to evaluating AI models in requirements engineering. We develop a web-based tool, followed by conducting experiments with four pre-trained large language models (LLMs). Subsequently, we will evaluate the satisfaction of user stories generation and prioritization through feedback from the system users.


Our study evaluates the effectiveness of advanced AI models—GPT-3.5, GPT-4, LLaMA3-70, and Mixtral-8B—in automating the requirements analysis process. The experiment captures detailed logs of each model's output, including the processing time, the similarity score, and the ranking of stories with different models. Moreover, we assess the quality of the user story against RE standards, such as the INVEST (Independent, Negotiable, Valuable, Estimable, Small, and Testable) quality framework \cite{INVEST} and ISO/IEC/IEEE 29148-2011 \cite{6146379}. We aim to determine each model's contribution to automating RE tasks and improving the quality of user story generation.

This study conducts experiments using a web tool environment that simulates real-world scenarios, providing relevant context for assessing the LLMs models' capabilities.


\subsection{Experiment design}

\begin{figure}
\includegraphics[width=\textwidth]{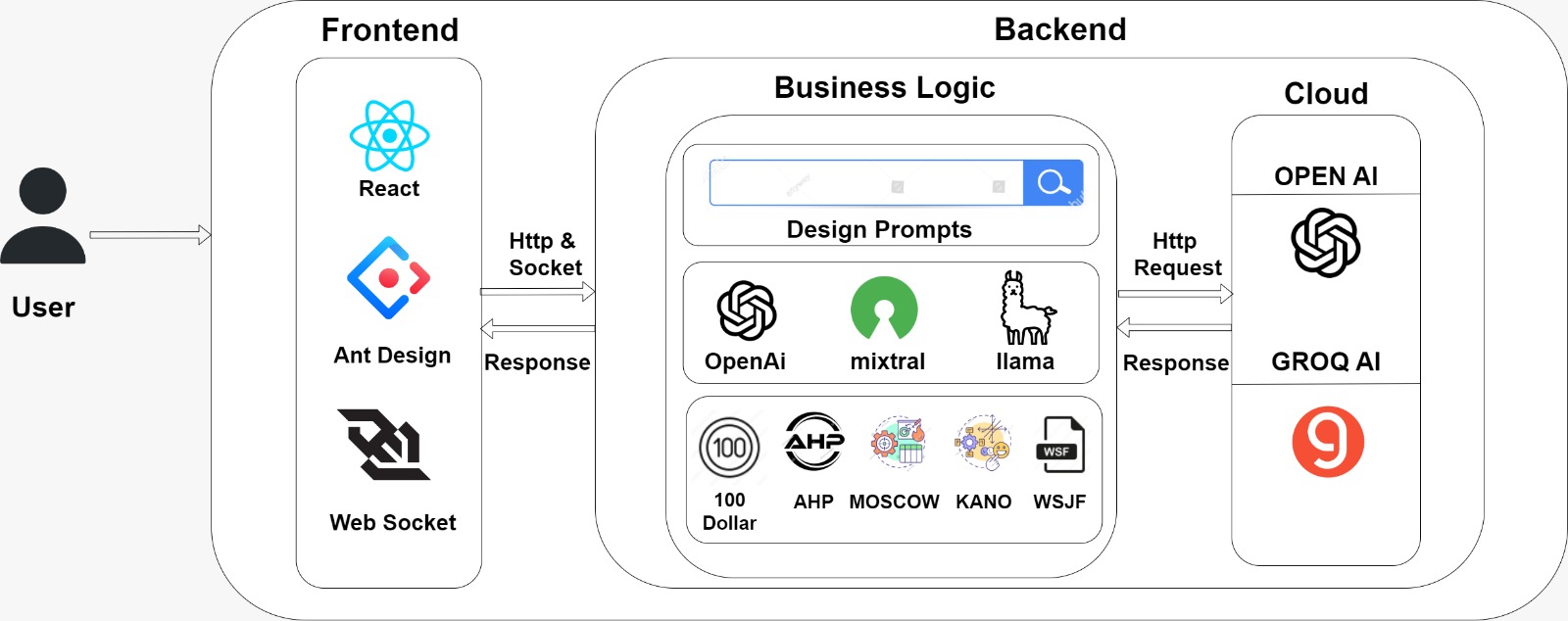}
\caption{Architecture of a Multi-Agent System}
\label{figure2}
\end{figure}

We designed a web-based application where the front end is built using HTML, CSS, JavaScript, React, and WebSocket. Real-time communication is managed via WebSocket, while other data is retrieved using HTTP requests. In the backend, we utilize Python and the Starlette framework. The backend receives HTTP calls from the client side and also calls LLM APIs from either OpenAI or GroqCloud. We use two API providers and four models to generate user stories, check their quality, and prioritize the work, as shown in Figure \ref{figure2}.

\textbf{User Story and Quality Framework}
  We follow the user story template and define criteria based on the \cite{openmobilityuserstory}. We check the compliance using the INVEST framework \cite{INVEST} and ISO/IEC/IEEE 29148-2011 standards \cite{6146379}.

\textbf{Prioritization Techniques}
 Three prioritization techniques are considered here. The Analytic Hierarchy Process (AHP) is a decision-making technique that uses pairwise comparisons and expert judgments to derive priority scales, integrating both subjective and objective aspects of decision-making \cite{somohano2021improving}. The Hundred Dollar Prioritization Method allows stakeholders to allocate 100 points to various requirements based on their importance \cite{khan2015comparison}. Weighted Shortest Job First (WSJF) prioritizes tasks by dividing the Cost of Delay (CoD) by the task duration, ensuring that high-value tasks are completed quickly, which aligns projects with key business objectives \cite{gopal2022requirement}.

 \subsection{Example projects} 
We chose four real project descriptions for empirical evaluation and feedback. From these project descriptions, we generate user stories, check the quality of the user story, and prioritize. The initial project description is as follows. 

\begin{tcolorbox}[colback=gray!5!white, colframe=gray!75!black, title=Project Descriptions, width=\textwidth]
\tiny 

\textbf{Project 1 (P1): FEMMa Oracle} \\The FEMMa Oracle project focuses on developing an AI assistant for the "Future Electrified Mobile Machines" (FEMMa) research project. The assistant will provide accurate responses about the project and feature a UI similar to ChatGPT with "FEMMa’s log." The project includes data collection from the FEMMa wiki, building a vector database, and developing both backend and frontend components. The prototype is set for early August 2024. Key aspects include project objectives, data collection methods, analysis, key findings, and practical applications. \textit{...for comprehensive details, refer to the full text.}

\textbf{Project 2 (P2): Supplier Data Management Platform} \\
This project addresses the challenge of managing and comparing data from over 500 vendors. The current system struggles with vendor diversity and cost tracking. The new platform will streamline data extraction from PDFs, translate and structure it for LLM use, and store it in MongoDB. The goal is to enable real-time analysis and provide insights into supplier offerings through a functional demo expected by late July. Use cases include ingredient and cost analysis for lasagna, supplier rate comparison for pasta, and recipe cost calculation. Future plans include API development for data uploads and enhanced analysis for better procurement and cost management. \textit{...for full information, see the detailed description.}

\textbf{Project 3 (P3): CV Analysis System} \\
This project aims to streamline the CV analysis process by developing a web application to parse and rank CVs based on job criteria. HR will set up campaigns, add job descriptions, and receive ranked CVs. The system involves user-uploaded CVs, parsing them based on specified criteria, and storing results in a vector-based database. An LLM chat assistant will assist with real-time queries. The system will identify top applicants and provide future real-time assistance reports. \textit{...for a detailed overview, see the full text.}

\textbf{Project 4 (P4): Tampere City RAG Application} \\
The Tampere City RAG (Retrieve and Generate) Application aims to develop an AI-driven chatbot to address queries about the Land Use and Construction Act in Finnish. Key data sources include Finnish legislation, city planning, and building permits websites. The approach involves web scraping, translating data for internal use, structuring data for an LLM, and storing it in a vector database like Pinecone or Qdrant. The chatbot will be deployed on the Tampere City website, handling questions such as "Can I cut down a tree from my plot?" and "What about a hedge? Need a permit?" An English version may be developed in the future. \textit{...for complete details, refer to the full text.}

\end{tcolorbox}

\subsection{Evaluation and Feedback}
We have three requirements analysis tasks: user story generation, quality enhancement, and prioritization. For user story generation, we evaluate word count, Application Programming Interface (API) response time, and semantic similarity score, and count distinct epics and stories from the tool. We determine the semantic similarity for each model by providing user stories and descriptions. Specifically, we employ the sentence transformer model \textbf{all-MiniLM-L6-v2} \cite{reimers2019sentence}, transforming sentences into embeddings or vectors that grasp their semantic meaning. 

The outputs of this model are sentence embeddings, which can be used for tasks such as semantic similarity, clustering, and classification. We calculate the score for how each user story semantically matches the project's description. The user stories that most closely fit the project details are then determined by looking at the mean scores from each model. We evaluate whether each story complies with the specified format for quality checks of user stories. This assessment determines if the story meets the given quality standards.

Regarding the evaluation of generated user story priority, we conduct a two-step process. First, we 
compare the ranking of user stories generated by different LLM models acting as agents, 
using various prioritization techniques.  
Next, we assess the consistency of these rankings to identify the optimal prioritization which is the ranking most frequently generated by the agents and prioritization techniques. These calculations are performed on a Google Colab CPU instance \footnote{https://colab.research.google.com/}. To visualize and compare these metrics effectively, a Python script was developed using pandas and matplotlib, resulting plots provide a side-by-side comparison of the model's performance across the projects. This enables us to display the outcomes for every project and identify 
the most techniques for prioritizing user stories.

In addition, we gather feedback from project teams on how satisfied they are with the requirements analysis results produced by the multi-agent system. The results are analyzed to determine the effectiveness of specific AI models in enhancing RE practices.

\section{Results}
\label{results}
\begin{figure}
\includegraphics[width=\textwidth]{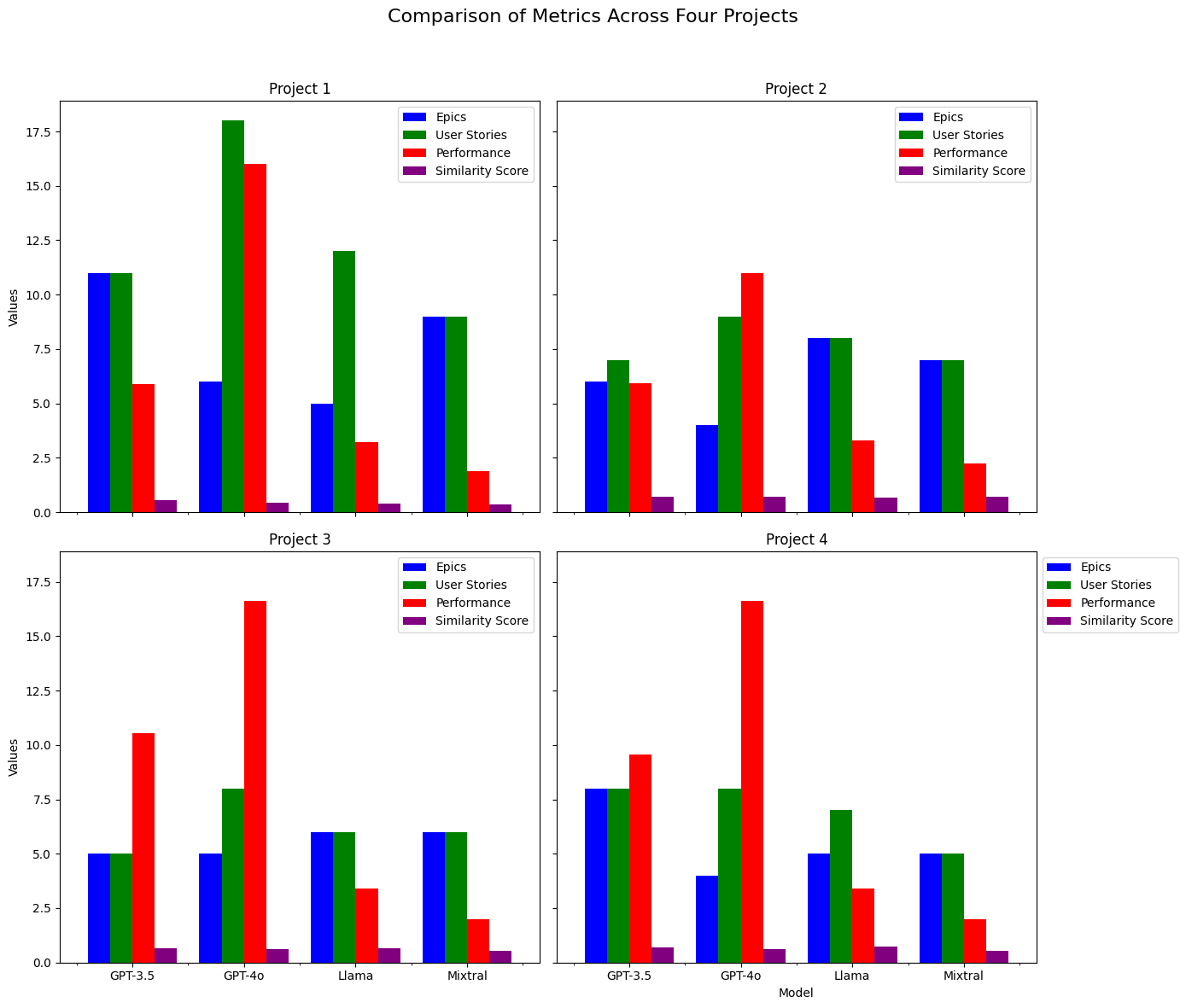}
\caption{Metrics Comparison for Four Projects Using Different Models.}
\label{figure4}
\end{figure}

We conducted experiments on four projects, 
converting project descriptions into good and priorities user stories. The results are presented with metrics such as 
the number of distinct stories, API response time, and the similarity score with the project description. Fig. \ref{figure4} displays these results for four projects, 
comparing the performance of four models, i.e. GPT-3.5, GPT-4o, Llama, and Mixtral, in analyzing requirements for each of the four projects.

In \textbf{Project 1}, GPT-3.5 produces 11 epics and 11 user stories, with an API response time of 5.90 seconds and a similarity score of 0.57. Comparatively, GPT-4o has a higher API response time of 16.00 seconds, despite a lower similarity score of 0.44. GPT-4o also generates more user stories (18) than epics (6). Llama and Mixtral generate five and nine epics, respectively, with lower API response times of 3.23 seconds and 1.88 seconds, and similarity scores of 0.38 and 0.36, respectively.

In \textbf{Project 2}, GPT-3.5 generates six epics and seven user stories, resulting in a high similarity score of 0.72 and an API response time of 5.91 seconds. In contrast, GPT-4o achieves a substantial API response time of 11.00 seconds and a similarity score of 0.71, while managing four unique epics and nine user stories. Llama and Mixtral demonstrate moderate performance, with API response times of 2.26 and 3.29 seconds, and similarity scores of 0.69 and 0.71, respectively, producing eight and seven epics each.

In \textbf{Project 3}, GPT-3.5 produces five epics and five user stories, with an API response time of 10.55 seconds and a similarity score of 0.65. GPT-4o continues to consume more time, with an API response time of 16.62 seconds and a similarity score of 0.62, producing five epics and eight user stories. Llama achieves an API response time of 3.41 seconds and a similarity score of 0.65, while Mixtral records an API response time of 1.99 seconds and a similarity score of 0.55, with both models generating six epics each. 

In \textbf{Project 4}, GPT-3.5 produces eight epics and eight user stories, with an API response time of 9.55 seconds and a similarity score of 0.71. GPT-4o generates four epics and eight user stories, with a long API response time of 16.62 seconds and a similarity score of 0.60. Llama and Mixtral each produce five epics, with Llama scoring 3.41 seconds for API response time and 0.72 for similarity, and Mixtral scoring 1.99 seconds for API response time and 0.55 for similarity.

As summary, GPT-4o shows higher API response times and also has high similarity scores. On the other hand, GPT-3.5 consistently maintains higher similarity scores than GPT-4o, while having lower API response times.

\subsection{Prioritization Results}
\begin{figure}
\includegraphics[width=\textwidth]{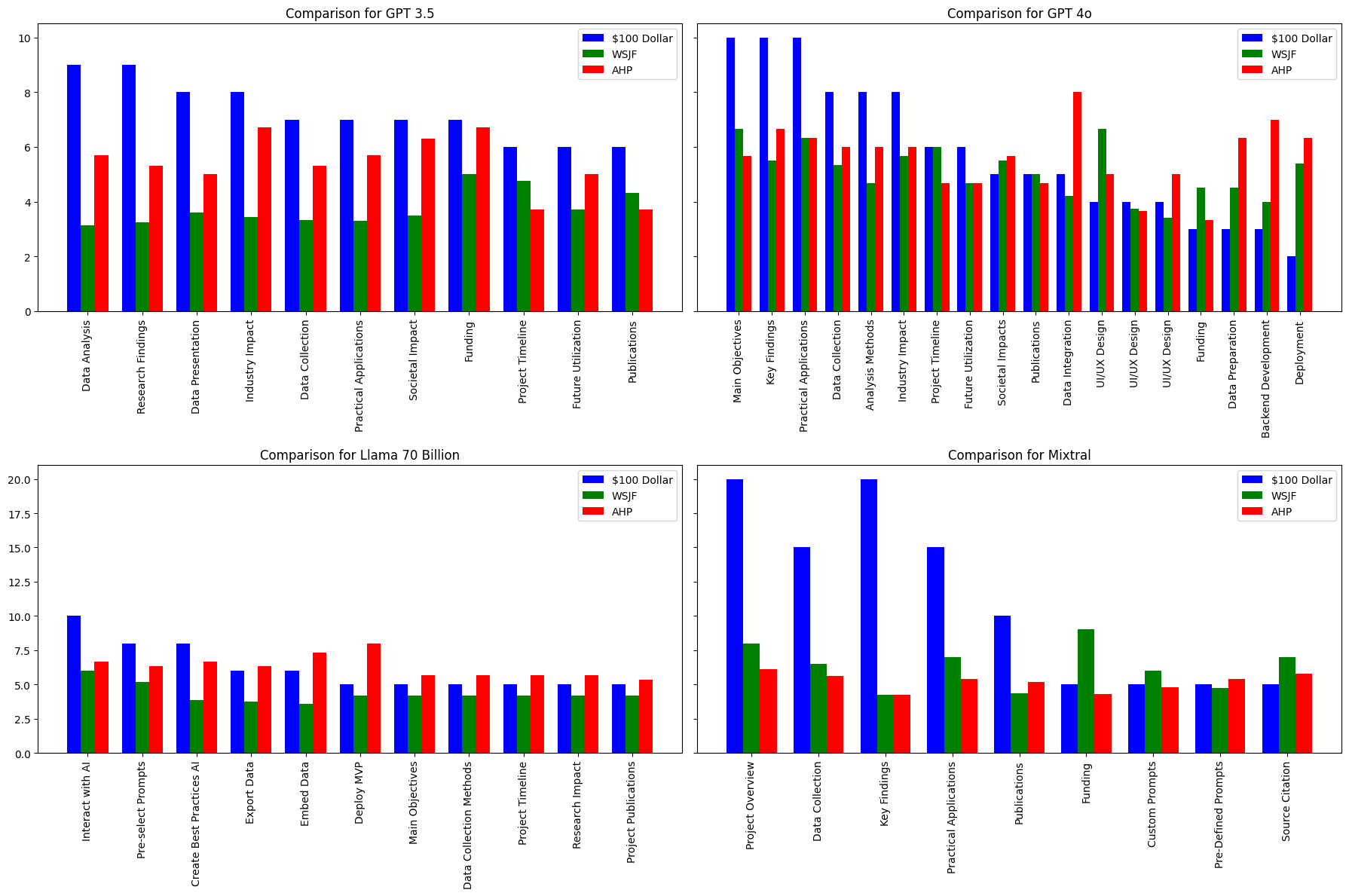}
\caption{Prioritization results of Project 1.}
\label{figure5}
\end{figure}

This study evaluates the prioritization of user stories on the basis of 
project descriptions using four models: GPT-3.5, GPT-4o, Llama 70 Billion, and Mixtral. We applied 
three prioritization techniques: 100 Dollar Allocation, Weighted Shortest Job First (WSJF), and Analytic Hierarchy Process (AHP). The results reveal distinct patterns for each technique and model, highlighting the relevance and consistency of user story prioritization. To illustrate these findings, We present the prioritization results for  Project 1, 
as shown in Fig. \ref{figure5}.

For \textbf{GPT-3.5}, Dollar Allocation provided the most consistent and relevant prioritization, with top-ranked user stories focusing on \textbf{Data Analysis} and \textbf{Research Findings} (both ranked 1.5). The WSJF score ranked \textbf{Funding} highest (ranked 1), emphasizing immediate objectives, while the AHP score highlighted \textbf{Industry Impact} and \textbf{Funding} (both ranked 1.5), emphasizing practical applications and industry relevance.

For \textbf{GPT-4o}, Dollar Allocation remained consistent with key objectives, ranking \textbf{Main Objectives}, \textbf{Key Findings}, and \textbf{Practical Applications} highest (all ranked 2). The WSJF score balanced interactions and design, with top priorities being \textbf{Main Objectives} and \textbf{UI/UX Design} (both ranked 1.5). The AHP score emphasized integration and development, prioritizing \textbf{Data Integration} (ranked 1) and \textbf{Backend Development} (ranked 2).

For \textbf{Llama 70 Billion}, Dollar Allocation focused on key interactions and prompt selections, ranking \textbf{Interact with AI} highest (ranked 1). The WSJF score balanced interactions and mid-tier tasks, with \textbf{Interact with AI} (ranked 1) as the top priority. The AHP score emphasized deployment and data embedding, with \textbf{Deploy MVP} (ranked 1) and \textbf{Embed Data} (ranked 2) as the highest priorities.

For \textbf{Mixtral}, Dollar Allocation prioritized an overview and key findings, ranking \textbf{Project Overview} and \textbf{Key Findings} highest (both ranked 1.5). The WSJF score emphasized funding and practical applications, with \textbf{Funding} (ranked 1) and \textbf{Project Overview} (ranked 2) as top priorities. The AHP score focused on the overview and data collection, with \textbf{Project Overview} (ranked 1) and \textbf{Source Citation} (ranked 2) highest.

Overall, \textbf{Dollar Allocation} generally provided more consistent and relevant prioritization across different models, emphasizing key objectives and practical applications. \textbf{WSJF and AHP scores} offered additional insights but with more variability, reflecting their sensitivity to different aspects of user stories and project requirements. The findings indicate that while Dollar Allocation is robust for strategic prioritization, WSJF and AHP provide valuable perspectives on immediate objectives and technical integration, respectively.

\begin{figure}
\includegraphics[width=\textwidth]{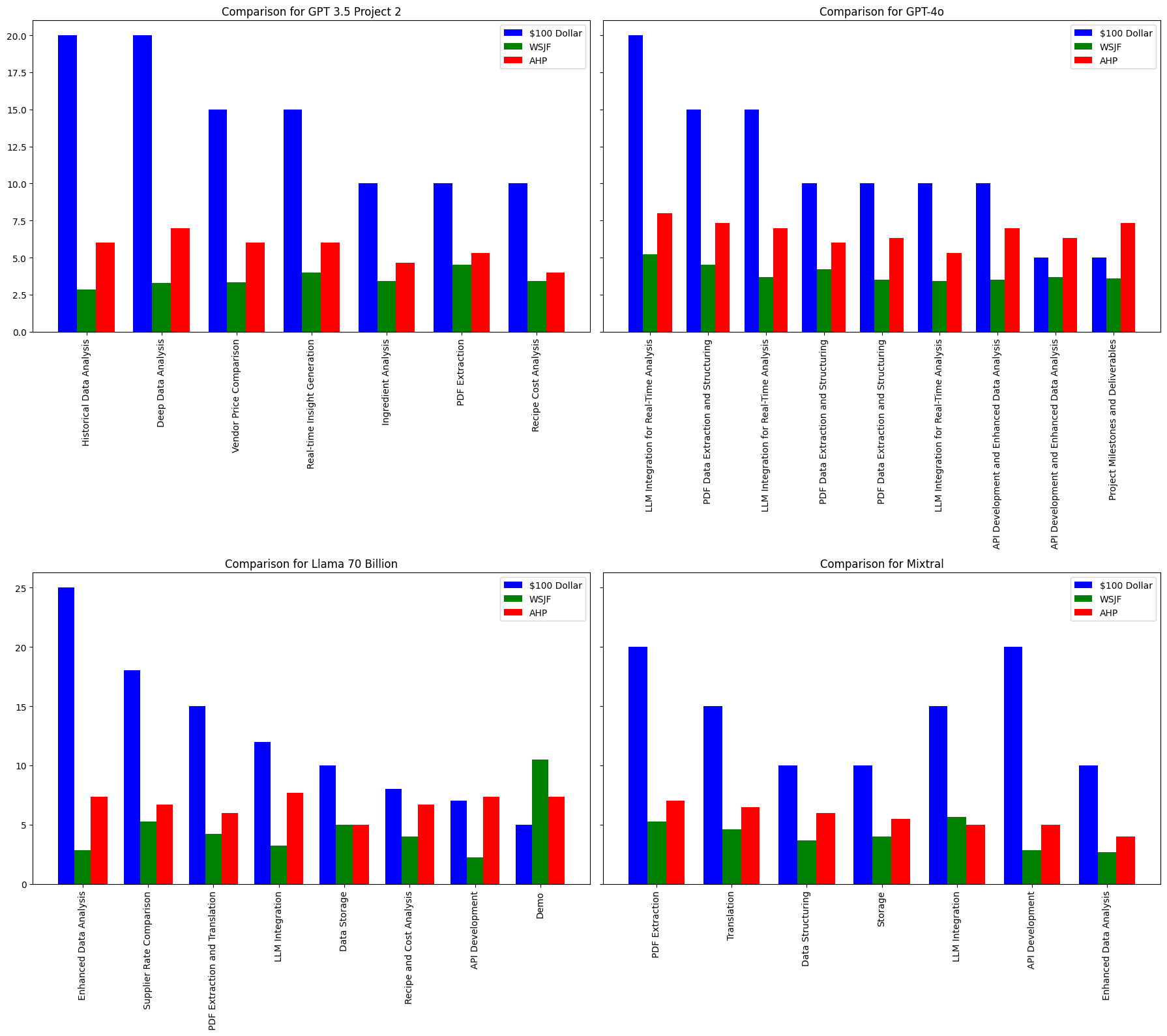}
\caption{Prioritization result of Project 2.}
\label{figure6}
\end{figure}

For Project 2, as shown in Figure \ref{figure6}, the results indicate that \textbf{GPT-3.5} scores high on these two stories 'Historical Data Analysis' and 'Deep Data Analysis'. \textbf{GPT-4o} emphasizes 'LLM Integration' and 'PDF Data Extraction', scoring the highest. \textbf{Llama 70 Billion} prioritizes 'Enhanced Data Analysis' and 'Supplier Rate Comparison' stories. \textbf{Mixtral} highlights and ranks 'PDF Extraction' and 'Translation', with high scores. Whereas, in  Project 1, the models prioritize differently. \textbf{GPT-3.5} balances 'Data Analysis' and 'Research Findings' stories. \textbf{GPT-4o} focuses on 'Main Objectives' and 'Key Findings' stories. \textbf{Llama 70 Billion} emphasizes 'Interact with AI' and 'Pre-select Prompts'. \textbf{Mixtral} highlights 'Project Overview' and 'Data Collection', achieving high scores.

\begin{table*}[ht]
\centering
\caption{Practitioners' Demographics and Their Assessment}
\label{tab:practitioners_assessment}
\resizebox{\textwidth}{!}{%
\begin{tabular}{|c|p{2.5cm}|c|c|p{3cm}|p{3cm}|}
\hline
\textbf{ID} & \textbf{Role of Practitioner} & \textbf{Experience} & \textbf{Satisfaction} & \textbf{Feedback} & \textbf{Suggestion} \\ \hline
P1 & Sr Developer & 5 Years & Good & Suggested enhancements in the user interface. & Include CRUD operations and remove hallucinations. \\ \hline
P2 & Product Owner & 8 Years & Satisfactory & Impressed with user story generation and prioritization. & Need acceptance criteria with more details. \\ \hline
P3 & Project Lead & 10 Years & Good & Some user stories missing technical steps. & Advised incorporating more technical steps. \\ \hline
P4 & Developer & 3 Years & Satisfactory & Suggested adding search criteria. & A database and sub-tasks could be added. \\ \hline
\end{tabular}
}
\end{table*}

In addition, we collected feedback and suggestions from practitioners on 
the outputs of the tool. A summary of their feedback is shown in Table \ref{tab:practitioners_assessment}. It shows that two participants indicated satisfaction with the generated outputs, while the other two expressed a good level of satisfaction, though they also suggested several improvements. These suggestions include adding functionality for adding, updating, and deleting records, providing more detailed acceptance criteria, and incorporating technical steps, and sub-tasks. Integrating a database to enhance search capabilities. Addressing hallucinations and ensuring consistency remain critical challenges. However, these recommendations underscore key areas where the system can be refined to better align with user needs.

\section{Discussion}
\label{discussion}
The findings of our study highlight the overall effectiveness and potential use of multi-agent systems and large language models in the field of software requirements engineering \cite{huang2023agentcoder}. Our approach simplifies the requirements analysis process and provides insights into how to generate and prioritize user stories across different projects using multi-agents. The results demonstrate that our method is flexible enough to handle various projects and shows great potential for increasing the productivity and precision of requirements engineering tasks. From table \ref{tab:practitioners_assessment}, feedback analysis shows that LLM-based models can serve as valuable tools to assist product owners during the early stages of requirements engineering. While agents can partially substitute for some roles with human involvement, integrating them as autonomous agents in the roles of product owners or requirements engineering agents requires further experimentation. Additionally, enhancing the performance of LLMs by mitigating hallucinations will improve the reliability and contextual accuracy of information during the requirements engineering phase.

Below, we explore the specifics of our findings for each research question.

\paragraph{AI agents assistance in requirements analysis (RQ1)}

In exploring this research question, we discovered that multi-agent systems can streamline and enhance the requirements analysis process. Our state-of-the-art approach employs four distinct AI agents. The Product Owner Agent is responsible for generating detailed user stories, facilitating stakeholder discussions, and evaluating the quality and compliance of these stories. The Quality Assurance Agent focuses on assessing the quality of user stories and participates in the prioritization of the stories. The Senior Developer Agent and QA understands project requirements and suggests prioritization based on the project description context. Lastly, the LLM Manager Agent takes results from all other agents and provides the final prioritization of user stories. This multiagent allows us to design prompts and integrate business logic to collaboratively parse, analyze, and display results effectively. This method may involve less time and effort than traditional approaches since it enables a more methodical and comprehensive needs analysis.

\paragraph{Effectiveness of Different Language Models in Performing Requirements Analysis Tasks (RQ2)} 
Using different prioritization techniques—\textbf{100 Dollar Allocation, WSJF, and AHP}—along with metrics such as the number of epics and user stories generated, API response time, and similarity scores, the performance of the four language models—GPT-3.5, GPT-4o, Llama 70 Billion, and Mixtral—was assessed across multiple projects.
\textbf{Epics and User Stories:} GPT-3.5 consistently generated a higher number of epics and maintained higher similarity scores compared to GPT-4o, indicating its robustness in complex project scenarios. However, GPT-4o produced more user stories overall but at the cost of longer API response times.\textbf{API Response Time:} Mixtral excelled with the quickest response times across all projects, making it particularly effective for tasks requiring rapid decision-making. This contrasts with GPT-4o’s longer response times, likely due to its extensive story generation.\textbf{Similarity Scores:} Llama 70 Billion and Mixtral achieved stable and moderate similarity scores, indicating their reliability in producing contextually relevant responses. Overall, GPT-3.5 performed consistently well across all projects.

\textbf{Prioritization Techniques:} The 100 Dollar Allocation technique consistently delivered strong prioritization results across different models, emphasizing key objectives and practical applications. WSJF and AHP provided additional insights but with more variability, reflecting their sensitivity to different aspects of the user stories.

Overall, the finding suggests that the 100 Dollar Allocation technique emerged as particularly effective in strategic prioritization, making it a valuable tool in guiding development focus in requirements analysis tasks. Each model displayed distinct strengths: Mixtral and Llama 70 Billion showed balanced performance across prioritization techniques, and GPT-3.5 reliably prioritized practical applications and data analysis, making it suitable for research and industry-related tasks. Integrating these models based on their strengths can optimize the requirements analysis process in different contexts.




As we discuss the strengths of our work, there are also limitations to our approach. Hallucination remains a significant challenge that could be mitigated by incorporating Retrieval-Augmented Generation (RAG) techniques. Feedback indicated that the tool would benefit from integrating a database and an edit system, which are currently absent. Additionally, adding more technical details to the user stories can improve prompt relevance and provide more context. Enhancing the profiles and roles of the agents will allow them to perform their tasks more efficiently. Further experimentation with multi-agent platforms and expanding the roles of agents in software requirements engineering tasks is necessary. Additionally, exploring other LLM models and fine-tuning them specifically for user story generation and prioritization is recommended. Integrating additional prioritization techniques could also offer deeper insights into the capabilities of LLMs.

\section{Conclusion}
\label{conclusion}
In this study, we developed a multi-agent tool that generates user stories, assesses their quality, and prioritizes them. We selected three prioritization techniques and employed four LLM models, conducting multiple experiments to evaluate their effectiveness. The use of four distinct AI agents—Product Owner Agent, Quality Assurance and Compliance Agent, Senior Developer Agent, and LLM Manager Agent—enabled us to generate user stories, facilitate stakeholder discussions, and evaluate quality and compliance. These agents also play a crucial role in the prioritization process.

We further assessed the performance of the four language models on requirements analysis tasks across multiple projects. GPT-3.5 consistently achieved higher similarity scores, while GPT-4o exhibited higher API response times but excelled in user story generation and prioritization. Llama and Mixtral were found to be faster in response time, though they had moderate similarity scores. Our findings indicate that multi-agent systems significantly enhance and streamline the requirements analysis process through the use of advanced AI agents.

The source code of the tool and the experimental results are available on GitHub\footnote{https://github.com/GPT-Laboratory/multiagent-prioritization}, encouraging further research and practical applications in this field. 

\section{Acknowledgments}

We express our sincere gratitude to Business Finland for their generous support and funding of our project. Their commitment to fostering innovation and supporting research initiatives has been instrumental in the success of our work.

\bibliographystyle{splncs04}
\bibliography{references}

\begin{thebibliography}{10}
\providecommand{\url}[1]{\texttt{#1}}
\providecommand{\urlprefix}{URL }
\providecommand{\doi}[1]{https://doi.org/#1}

\bibitem{INVEST}
Wake, b: Invest in good stories, and smart tasks. \url{https://xp123.com/articles/invest-in-good-stories-and-smart-tasks/} (Accessed: 2024-1-10)

\bibitem{userstoriejune}
Abed, O., Nebe, K., Abdellatif, A.B.: Ai-generated user stories supporting human-centred development: An investigation on quality. In: Stephanidis, C., Antona, M., Ntoa, S., Salvendy, G. (eds.) HCI International 2024 Posters. pp. 3--13. Springer Nature Switzerland, Cham (2024)

\bibitem{alawaji2024evaluating}
Alawaji, B., Hakami, M., Alshemaimri, B.: Evaluating generative language models with prompt engineering for categorizing user stories to its sector domains. In: 2024 IEEE 9th International Conference for Convergence in Technology (I2CT). pp.~1--8. IEEE (2024)

\bibitem{dong2023self}
Dong, Y., Jiang, X., Jin, Z., Li, G.: Self-collaboration code generation via chatgpt. arXiv preprint arXiv:2304.07590  (2023)

\bibitem{feng2024prompting}
Feng, S., Chen, C.: Prompting is all you need: Automated android bug replay with large language models. In: Proceedings of the 46th IEEE/ACM International Conference on Software Engineering. pp. 1--13 (2024)

\bibitem{feng2023investigating}
Feng, Y., Vanam, S., Cherukupally, M., Zheng, W., Qiu, M., Chen, H.: Investigating code generation performance of chat-gpt with crowdsourcing social data. In: Proceedings of the 47th IEEE Computer Software and Applications Conference. pp. 1--10 (2023)

\bibitem{gopal2022requirement}
Gopal, M., Yacoob, A.O.: Requirement engineering using scaledagile framework{\textregistered}(safe) in automotiveindustry: Practices and challenges (2022)

\bibitem{herwanto2024automating}
Herwanto, G.B.: Automating data flow diagram generation from user stories using large language models. In: 7th Workshop on Natural Language Processing for Requirements Engineering (2024)

\bibitem{huang2023agentcoder}
Huang, D., Bu, Q., Zhang, J.M., Luck, M., Cui, H.: Agentcoder: Multi-agent-based code generation with iterative testing and optimisation. arXiv preprint arXiv:2312.13010  (2023)

\bibitem{6146379}
{ISO/IEC/IEEE International Standard 29148:2011(E)}: Iso/iec/ieee international standard - systems and software engineering -- life cycle processes -- requirements engineering. ISO/IEC/IEEE 29148:2011(E) pp. 1--94 (2011). \doi{10.1109/IEEESTD.2011.6146379}

\bibitem{khan2015comparison}
Khan, J.A., Rehman, I.U., Khan, Y.H., Khan, I.J., Rashid, S.: Comparison of requirement prioritization techniques to find best prioritization technique. International Journal of Modern Education and Computer Science  \textbf{7}(11), ~53 (2015)

\bibitem{ma2023scope}
Ma, W., Liu, S., Wang, W., Hu, Q., Liu, Y., Zhang, C., Nie, L., Liu, Y.: The scope of chatgpt in software engineering: A thorough investigation. arXiv preprint arXiv:2305.12138  (2023)

\bibitem{nguyen2023generative}
Nguyen-Duc, A., Cabrero-Daniel, B., Przybylek, A., Arora, C., Khanna, D., Herda, T., Rafiq, U., Melegati, J., Guerra, E., Kemell, K.K., et~al.: Generative artificial intelligence for software engineering--a research agenda. arXiv preprint arXiv:2310.18648  (2023)

\bibitem{rajbhoj2024accelerating}
Rajbhoj, A., Somase, A., Kulkarni, P., Kulkarni, V.: Accelerating software development using generative ai: Chatgpt case study. In: Proceedings of the 17th Innovations in Software Engineering Conference. pp. 1--11 (2024)

\bibitem{rasheed2024codepori}
Rasheed, Z., Waseem, M., Saari, M., Syst{\"a}, K., Abrahamsson, P.: Codepori: Large scale model for autonomous software development by using multi-agents. arXiv preprint arXiv:2402.01411  (2024)

\bibitem{reimers2019sentence}
Reimers, N., Gurevych, I.: Sentence-bert: Sentence embeddings using siamese bert-networks. arXiv preprint arXiv:1908.10084  (2019)

\bibitem{runeson2009guidelines}
Runeson, P., H{\"o}st, M.: Guidelines for conducting and reporting case study research in software engineering. Empirical Software Engineering  \textbf{14}(2), ~131 (2009)

\bibitem{sami2024system}
Sami, A.M., Rasheed, Z., Kemell, K.K., Waseem, M., Kilamo, T., Saari, M., Duc, A.N., Syst{\"a}, K., Abrahamsson, P.: System for systematic literature review using multiple ai agents: Concept and an empirical evaluation. arXiv preprint arXiv:2403.08399  (2024)

\bibitem{sami2024prioritizing}
Sami, M.A., Rasheed, Z., Waseem, M., Zhang, Z., Herda, T., Abrahamsson, P.: Prioritizing software requirements using large language models (2024)

\bibitem{sauvola2024future}
Sauvola, J., Tarkoma, S., Klemettinen, M., Riekki, J., Doermann, D.: Future of software development with generative ai. Automated Software Engineering  \textbf{31}(1), ~26 (2024)

\bibitem{somohano2021improving}
Somohano-Murrieta, J.C.B., Ochar{\'a}n-Hern{\'a}ndez, J.O., S{\'a}nchez-Garc{\'\i}a, {\'A}.J., Lim{\'o}n, X., los {\'A}ngeles Arenas-Vald{\'e}s, M.d.: Improving the analytic hierarchy process for requirements prioritization using evolutionary computing. Programming and Computer Software  \textbf{47},  746--756 (2021)

\bibitem{treude2023navigating}
Treude, C.: Navigating complexity in software engineering: A prototype for comparing gpt-n solutions. arXiv preprint arXiv:2301.12169  (2023)

\bibitem{waseem2023artificial}
Waseem, M., Das, T., Paloniemi, T., Koivisto, M., R{\"a}s{\"a}nen, E., Set{\"a}l{\"a}, M., Mikkonen, T.: Artificial intelligence procurement assistant: Enhancing bid evaluation. In: International Conference on Software Business. pp. 108--114. Springer Nature Switzerland Cham (2023)

\bibitem{wohlin2012experimentation}
Wohlin, C., Runeson, P., H{\"o}st, M., Ohlsson, M.C., Regnell, B., Wessl{\'e}n, A.: Experimentation in Software Engineering. Springer Science \& Business Media (2012)

\bibitem{xie2023chatunitest}
Xie, Z., Chen, Y., Zhi, C., Deng, S., Yin, J.: Chatunitest: a chatgpt-based automated unit test generation tool. arXiv preprint arXiv:2305.04764  (2023)

\bibitem{zzyllm}
Zhang, Z., Rayhan, M., Herda, T., Goisauf, M., Abrahamsson, P.: Llm-based agents for automating the enhancement of user story quality: An early report. In: International Conference on Agile Software Development. pp. 117--126. Springer Nature Switzerland Cham (2024)

\bibitem{zhong2024ldb}
Zhong, L., Wang, Z., Shang, J.: Ldb: A large language model debugger via verifying runtime execution step-by-step. arXiv preprint arXiv:2402.16906  (2024)

\bibitem{zhou2023language}
Zhou, A., Yan, K., Shlapentokh-Rothman, M., Wang, H., Wang, Y.X.: Language agent tree search unifies reasoning acting and planning in language models. arXiv preprint arXiv:2310.04406  (2023)

\end{thebibliography}

\end{document}